\documentclass[10pt,twocolumn,twoside]{IEEEtran}
\usepackage{graphicx}
\usepackage{amsmath}

\usepackage{amsfonts}
\usepackage{amssymb}
\usepackage{array}
\newcommand{\PreserveBackslash}[1]{\let\temp=\\#1\let\\=\temp}
\newcolumntype{C}[1]{>{\PreserveBackslash\centering}p{#1}}
\newcolumntype{R}[1]{>{\PreserveBackslash\raggedleft}p{#1}}
\newcolumntype{L}[1]{>{\PreserveBackslash\raggedright}p{#1}}
\usepackage[usenames]{color}
\usepackage{colortbl,booktabs}
\usepackage{tabularx}
\usepackage{multicol}
\usepackage{booktabs}
\usepackage[Symbol]{upgreek}
\usepackage{subfigure}
\usepackage{bm}
\usepackage{cite}
\usepackage{balance}
\usepackage[ruled,vlined]{algorithm2e}

\usepackage{threeparttable}
\usepackage{amsmath}
\usepackage{dcolumn}
\usepackage{multirow}
\usepackage{amsfonts}

\newcolumntype{d}[1]{D{.}{.}{#1}}

\begin{document}
	
	\bibliographystyle{IEEEtran} 
	\title{Empowering Near-Field Communications in Low-Altitude Economy with LLM: Fundamentals, Potentials, Solutions, and Future Directions}
	
	\author {{Zhuo Xu, Tianyue Zheng, and Linglong Dai,~\IEEEmembership{Fellow,~IEEE}}
 
		\thanks{The authors are with the Department of Electronic Engineering, Tsinghua University as well as the State Key Laboratory of Space Network and Communications, Beijing 100084, China (e-mails: \{xz23, zhengty22\}@mails.tsinghua.edu.cn,  daill@tsinghua.edu.cn).}
	}
	
	\maketitle
	\vspace{-0mm}
	\begin{abstract}

The low-altitude economy (LAE) is gaining significant attention from academia and industry. Fortunately, LAE naturally aligns with near-field communications in extremely large-scale MIMO (XL-MIMO) systems. By leveraging near-field beamfocusing, LAE can precisely direct beam energy to unmanned aerial vehicles, while the additional distance dimension boosts overall spectrum efficiency. However, near-field communications in LAE still face several challenges, such as the increase in signal processing complexity and the necessity of distinguishing between far and near-field users. Inspired by the large language models (LLM) with powerful ability to handle complex problems, we apply LLM to solve challenges of near-field communications in LAE. The objective of this article is to provide a comprehensive analysis and discussion on LLM-empowered near-field communications in LAE. Specifically, we first introduce fundamentals of LLM and near-field communications, including the key advantages of LLM and key characteristics of near-field communications. Then, we reveal the opportunities and challenges of near-field communications in LAE. To address these challenges, we present a LLM-based scheme for near-field communications in LAE, and provide a case study which jointly distinguishes far and near-field users and designs multi-user precoding matrix. Finally, we outline and highlight several future research directions and open issues.

	\end{abstract}
	
	
\section{Introduction}

Recently, the low-altitude economy (LAE) has garnered substantial interest from both industry and academia across multiple countries\cite{wan2024sensing,tang2024cooperative}. This emerging sector leverages various flying platforms, notably unmanned aerial vehicles (UAVs), to support diverse applications such as urban transportation, logistics, agriculture, and tourism. From a wireless communications perspective, LAE networks utilize UAVs to fulfill diverse communication needs, capitalizing on the airspace’s greater mobility compared to terrestrial networks. Ensuring the stable and secure operation of UAVs is critical, requiring seamless wireless connectivity and precise trajectory planning and tracking to support these dynamic operations\cite{zeng2019accessing}.

To enable the successful deployment of LAE, extremely large-scale MIMO (XL-MIMO) has emerged as a promising technology. Unlike traditional massive MIMO, XL-MIMO employs extremely large-scale antenna arrays (ELAA), offering superior spatial resolution and multiplexing gains\cite{wang2024tutorial,lu2024tutorial}. As the number of antennas at the base station (BS) increases, the near-field region expands, necessitating the use of a spherical-wave model over the conventional planar-wave mode. This near-field channel, influenced by both angle and distance, provides enhanced focusing capabilities, concentrating beam energy at specific locations like a targeted spotlight, thus unlocking significant potential for XL-MIMO systems\cite{zhang2022beam}.

Fortunately, the synergy between LAE and near-field communications in XL-MIMO systems is a natural fit. Unlike ground users, UAVs, operating at practical altitudes, are more likely to fall within the near-field region, allowing them to fully leverage near-field benefits. Specifically, LAE can harness the near-field beamfocusing feature to precisely direct energy to individual UAV positions, reducing interference. Moreover, beyond the traditional far-field spatial division multiple access (SDMA), the near-field location division multiple access (LDMA) enables simultaneous service to UAVs at the same angle but different distances, boosting overall spectrum efficiency through an additional distance dimension.

However, near-field communications in LAE still face several challenges, including increased signal processing complexity and the need for accurate user classification. Inspired by the powerful advantages of LLM such as their adaptability to dynamic environments, strong generalization capabilities, and ability to handle complex data patterns\cite{jiang2024large,jiang2025comprehensive,guo2025prompt}, we propose leveraging LLM to address these challenges, optimizing near-field communication performance in LAE networks.

In this article, we systematically investigate and summarize LLM-empowered near-field communication in LAE. The principal contributions of this article can be briefly summarized as follows:

\begin{itemize}
    \item To begin with, the fundamentals of LLM and near-field communications are introduced, laying the groundwork for their application in LAE. Specifically, we explore the key advantages of LLM, including their adaptability and data processing capabilities, which make them highly effective for addressing complex challenges in wireless communication systems. Additionally, we highlight the essential characteristics of near-field communications, such as precise beamfocusing, which is particularly suited to the dynamic needs of LAE environments.

   \item Additionally, we examine the promising opportunities arising from the seamless integration of LLM, near-field communications, and LAE. However, we also identify critical challenges, including the increased signal processing complexity due to the added distance dimension and the difficulty in distinguishing between far-field and near-field users, both of which require innovative solutions to ensure practical deployment.

   \item Moreover, we propose a novel LLM-based scheme tailored for near-field communications in LAE, offering a detailed analysis of its overall framework and core principles. This approach leverages LLMs to enhance performance in various tasks, including user classification and multi-user precoding. To validate its effectiveness, we present a specific case study and provide simulation results, demonstrating significant improvements in accuracy and efficiency, thus paving the way for advanced LAE networking solutions.

   \item Finally, several future research directions and open issues are provided, including the development of LLM-empowered multi-task networks for LAE, the exploration of multi-modal LLMs tailored for LAE applications, and the deployment and verification of LLMs to ensure robust performance in real-world LAE environments.

\end{itemize}


\section{Fundamentals of LLM and near-field communications}\label{S2}
In this section, the fundamentals and key advantages of LLM are first illustrated. Then, key characteristics of near-field communications are provided and discussed.

\subsection{Key Advantages of LLM}\label{II-A}

LLMs, such as the GPT series, are advanced artificial intelligence systems built on transformer architectures, a type of neural network designed to process sequential data effectively. These models are trained on massive datasets, often comprising billions of words from diverse sources, enabling them to understand and generate human-like text. The transformer architecture relies on mechanisms like self-attention to capture long-range dependencies in data, making LLMs highly effective for tasks requiring complex pattern recognition. In the context of the LAE, LLM can be adapted to process wireless communication data, such as channel state information, to optimize tasks like user classification and signal resource allocation.

\begin{figure}[t!]
	\centering
	\includegraphics[width=0.49\textwidth]{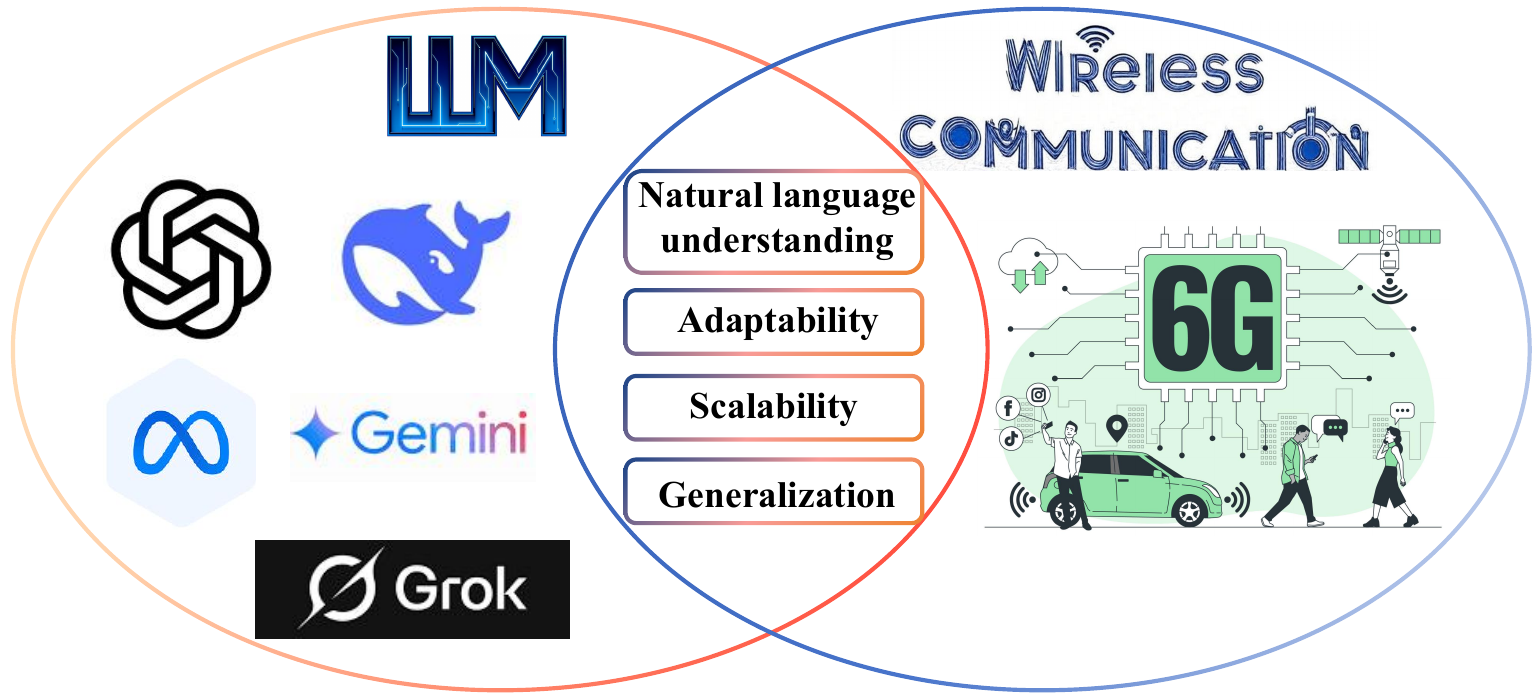}
	\caption{Key advantages of applying LLM in wireless communication systems.}
	\label{figurellmadvantage}
\end{figure}

With powerful reasoning and inference capabilities, LLM excels to tackle the complex problems, leveraging their scalability and adaptability to achieve superior performance. As illustrated in Fig.~\ref{figurellmadvantage}, the key advantages of applying LLM to solve problems in wireless communication systems can be summarized as:





\textbf{\textit{1) Natural language understanding:}} LLM excels at interpreting complex data patterns, a capability that extends beyond text to numerical data like wireless channel information. This enables LLMs to analyze intricate relationships within communication systems and understand user intents and needs, which is critical for near-field communications in LAE.

\textbf{\textit{2) Adaptability:}} Pre-trained on diverse datasets, LLM  possess inherent few-shot/zero-shot learning capabilities. 
Thus, through fine-tuning , LLM exhibit excellent adaptability and can be customized for specific tasks without retraining the entire model. For instance, a pretrained GPT-2 model can be adjusted to classify near-field versus far-field users or optimize signal directions, adapting to the dynamic conditions of LAE environments where UAVs operate in rapidly changing urban airspace.

\textbf{\textit{3) Feature extraction ability:}} Equipped with a large amount of parameters, LLM are designed to handle large-scale, high-dimensional data and address complex optimization problem. In XL-MIMO systems, which involve numerous antennas and users, LLMs can  efficiently extract features from high-dimensional complex data, ensuring performance remains robust as the network scales.

\textbf{\textit{4) Generalization:}} Compared to traditional methods, LLM exhibit strong generalization, performing reliably across varied scenarios without extensive retraining. This is particularly valuable in LAE, where conditions like user density, weather, or interference can vary, requiring adaptable communication solutions.

\subsection{Key Characteristics of Near-Field Communications}

As shown in Fig.~\ref{figurenearfieldcommunications}, the fundamental differences between near-field communications and traditional far-field communications lies in their distinct channel characteristics and the corresponding transmission designs. The classical far-field communications involve relatively long propagation distances, where electromagnetic waves can be reliably approximated by planar waves, rendering wavefront curvature negligible and relying primarily on angular domain information. By contrast, in near-field communications, signals are transmitted over shorter distances, typically within the Rayleigh distance, where electromagnetic waves exhibit spherical wavefront characteristics. This results in significant wavefront curvature, with signal phase and amplitude varying intricately as functions of both distance and angle, necessitating joint consideration of the angular and distance domains in the space. This disparity in propagation models leads to substantial differences in the corresponding transmission designs (such as including channel modeling, channel estimation, beamforming, precoding, etc.) between far-field and near-field communications. Particularly, far-field communications usually exploit directional beams in the angular domain, while near-field communications leverage the spherical wave to enable two-dimensional resolution in the “angle-distance” domain. 

The key characteristics of near-field communications different from far-field communications can be summarized mainly lie in the following three aspects:

 	\begin{figure*}
	 	\begin{center}
	 		\vspace*{-1mm}\hspace*{-1mm}\includegraphics[width=0.85
    \linewidth]{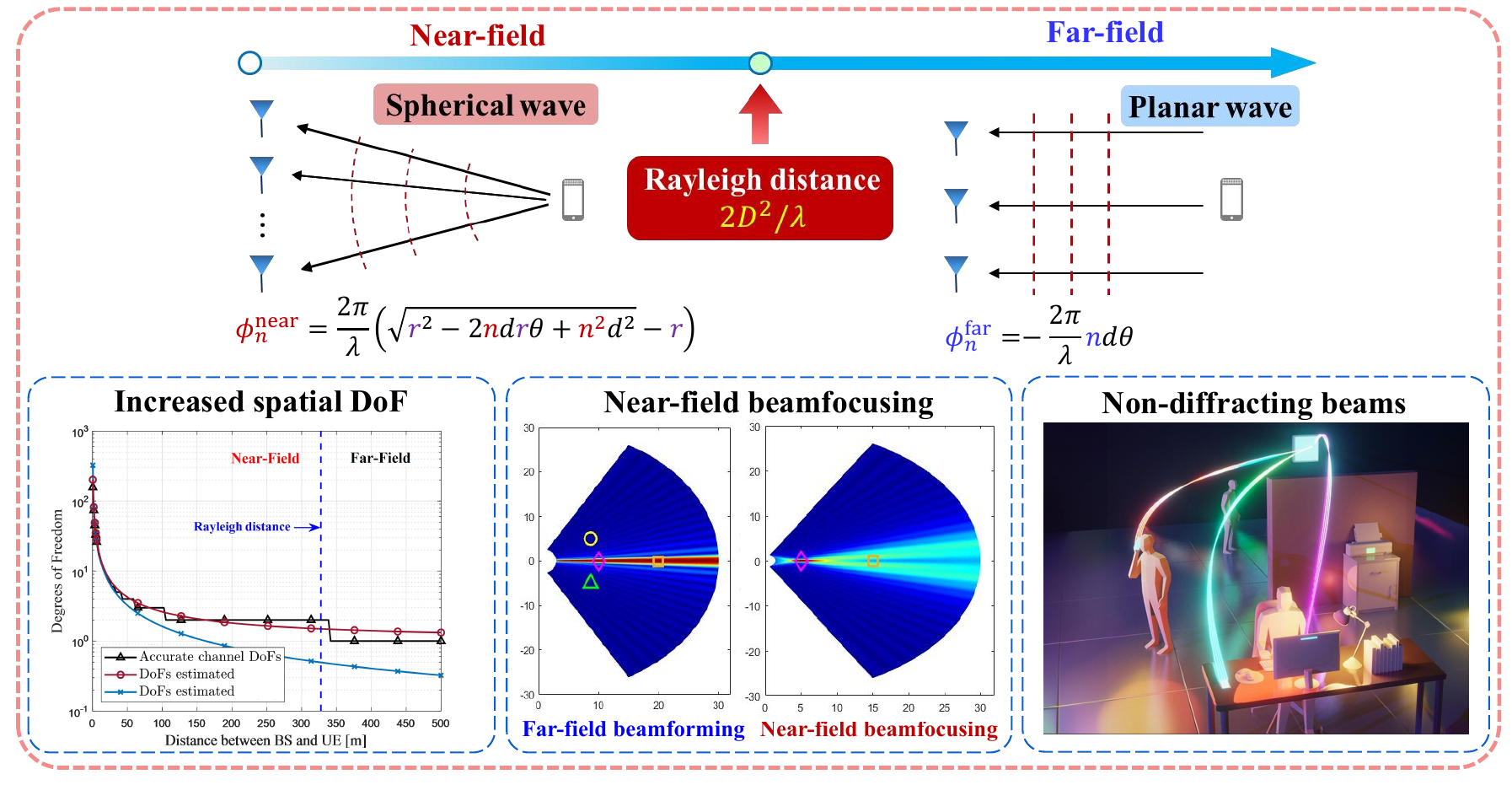}
	 	\end{center}\vspace*{-2mm}	
		\centering
		\caption{Key characteristics of near-field communications.} 
		\vspace*{-2mm}
		\label{figurenearfieldcommunications} 	
	 \end{figure*}

\textbf{\textit{1) Increased spatial degrees of freedom:}} Near-field communications capitalize on the spherical wave to unlock the additional degrees of freedom (DoF) in the distance domain. For example, the spatial DoF of far-field line-of-sight channels remains only one no matter how many antennas are used, while the spatial DoF of near-field line-of-sight channels will be significantly increased. This increased DoF can be effectively exploited to improve the spectral efficiency in single-user systems.

\textbf{\textit{2) Near-field beamfocusing:}} As shown in Fig.~\ref{figurenearfieldcommunications}, different from classical far-field beamforming only focusing its energy on specific angles, near-field communications facilitate precise beam focusing, concentrating signal energy at targeted spatial locations determined by angles and distances simultaneously. The near-field beamfocusing is just like a controllable convex lens, which focus most of the energy on the focal point while little power is leaked outside the focal point. This unique feature in the two-dimensional “angle-distance” domain can be directly used to enhance the spectral efficiency, increase the sensing accuracy, and improve the energy efficiency, etc.

\textbf{\textit{3) Non-diffracting beams:}} Additionally, near-field propagation enables the design of innovative beamforming techniques, e.g., one can converge the radiated electromagnetic wave energy along a strip to maintain diffraction-free propagation, which is known as non-diffracting beams (including Bessel beam and Airy beam). Non-diffracting beam avoids the limited energy convergence area associated with near-field focusing, allowing the signal to be maintained over longer distances and effectively extending the transmission range. It also offers new features, such as self-healing, enabling transmission around obstacles, making it a promising technology for future 6G new applications.

So far, we have discussed the key advantages of LLM and key characteristics of near-field communications. The opportunities and challenges of near-field communications in LAE will be analyzed in the following section.

\section{Opportunities and Challenges of near-field communications in LAE}

In this section, the opportunities and challenges of near-field communications in LAE are provided in detail.

\subsection{Opportunities}

\textit{\textbf{1) Opportunities for Seamless Integration of LAE and Near-Field Communications:}} It should be noted that in the future XL-MIMO systems, the LAE and near-field communications can align seamlessly. To ensure the successful implementation of LAE, the stable and safe operation of UAVs is particularly important. Specifically, the UAVs require seamless wireless communication connections and accurate trajectory planning and tracking. Fortunately, the LAE network can utilize the near-field beamfocusing characteristic to accurately focus the beam energy to the positions of different UAVs, boosting their spectrum efficiency and localization accuracy in the crowded urban skies. Beside, compared with users on the ground, UAVs are at a more practical height relative to the BS antenna arrays, which makes them more likely to be situated within the near-field region and consequently reap the benefits of near-field communications.

\textit{\textbf{2) Opportunities for Applying LLM into Near-Field Communications in LAE:}} Thanks to the key advantages of LLM discussed in subsection \ref{II-A}, there will be several opportunities to apply LLM into near-field communications in LAE. Specifically, LLM can harness their adaptability and scalability to optimize near-field beamfocusing, enabling precise signal delivery to UAVs in dynamic urban environments. This enhances drone coordination by quickly classifying users and adjusting power allocation in real time, ensuring seamless communication even as UAV traffic grows. Additionally, LLM’ ability to generalize across diverse scenarios allows them to handle unpredictable conditions like interference or noise, improving network reliability. By integrating LLMs, LAE networks can achieve greater intelligence, supporting innovative applications such as autonomous delivery systems and real-time air traffic management, thus driving the future of aerial connectivity.

\subsection{Challenges}

Although LLM-empowered near-field communications in LAE brings some new opportunities, it also faces several challenges. The main challenges can be summarized as follows:

\textbf{\textit{1) The Challenge of Signal Processing Complexity:}} One major challenge is the increased signal processing complexity introduced by the additional distance dimension in near-field communications. Unlike traditional far-field communications, which focus solely on angular beamsteering, near-field systems must account for both angle and distance, significantly complicating the processing demands. This complexity is particularly pronounced in LAE scenarios, where base station antenna tilt plays a critical role in performance. To maintain optimal connectivity with fast-moving UAVs, the antenna tilt may need real-time adjustments, adding further computational strain and requiring advanced optimization techniques.

\textbf{\textit{2) The Challenge of Distinguishing between Far and Near-Field Users:}} Another critical challenge lies in distinguishing between far and near-field users. Misclassifying all users as far-field can lead to significant performance losses, as the far-field planar-wave model fails to accurately represent the near-field region, reducing spectrum efficiency. Conversely, treating all users as near-field users introduces unnecessary computational and storage overhead due to the extra distance dimension, making it unacceptable for practical systems. Besides, due to the difficulty in obtaining the precise user distance, it is infeasible to distinguish them by directly comparing their distance with the Rayleigh distance. Therefore, how to achieve precise and rapid differentiation between far and near-field users becomes particularly crucial\cite{zhang2023mixed}.


So far, we have discussed the opportunities and challenges of near-field communications in LAE. In the following section, we will propose and analyze our LLM-based solutions.

\section{LLM-based Solutions for Near-Field Communications in LAE}

In this section, the overall framework and core principles of proposed LLM-based solutions is first discussed. Then, we present specific cases and provide simulation results to demonstrate its superior performance.

 	\begin{figure*}
	 	\begin{center}
	 		\vspace*{-1mm}\hspace*{-1mm}\includegraphics[width=0.95
    \linewidth]{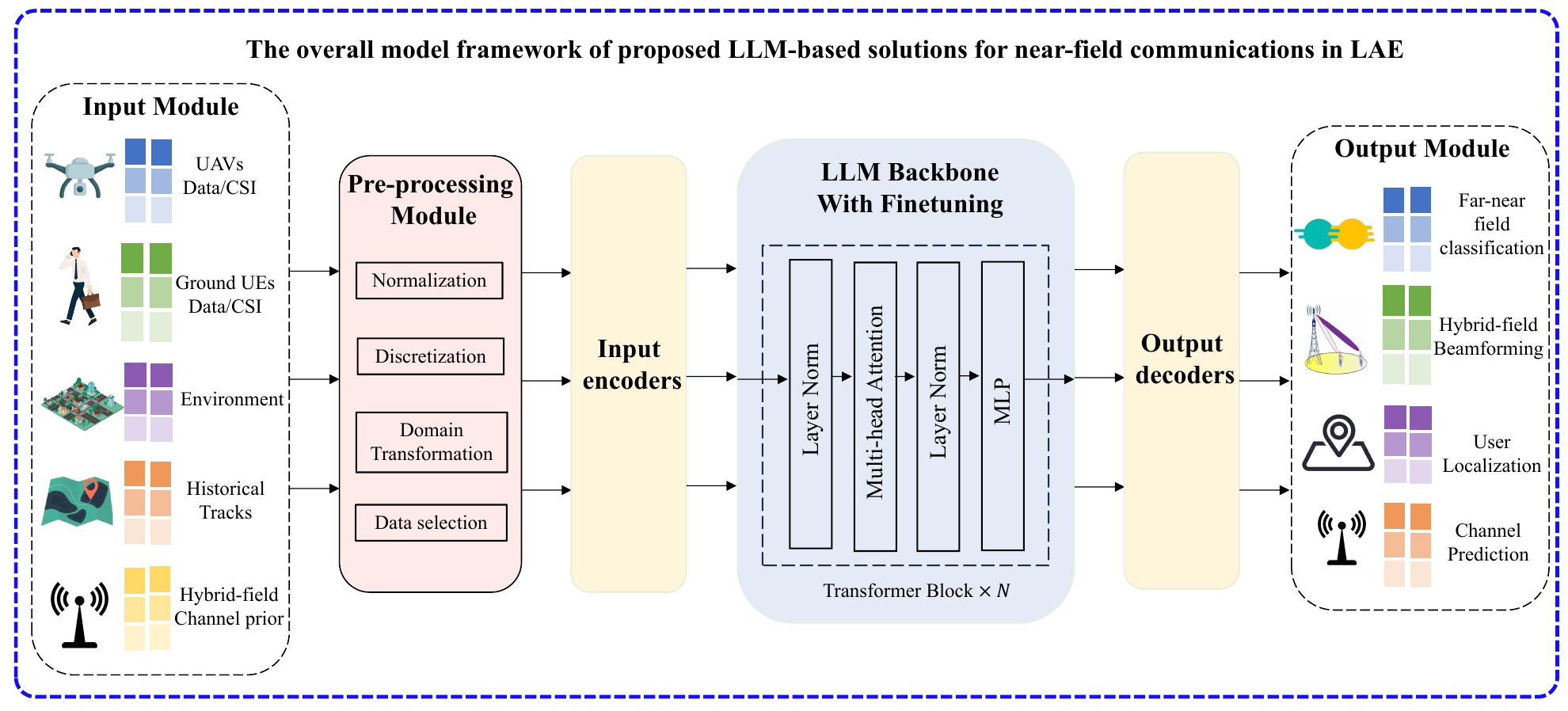}
	 	\end{center}\vspace*{-2mm}	
		\centering
		\caption{The overall model framework of proposed LLM-based solutions for near-field communications in LAE.} 
		\vspace*{-2mm}
		\label{figureoverallframework} 	
	 \end{figure*}

\subsection{Overall Framework and Core Principles}

To address the above challenges for near-field communications in LAE, we first apply LLM and give a overall framework of proposed LLM-based solutions. Specifically, the overall framework of proposed LLM-based solutions is illustrated in Fig.~\ref{figureoverallframework}, including \textit{Input Module}, \textit{Pre-processing Module}, \textit{Input encoders}, \textit{LLM Backbone with Finetuning}, \textit{Output decoders}, and \textit{Output Module}.

\noindent \textbf{\textit{1): Input Module:}}

The \textit{Input Module} serves as the entry point for data collection, which aims to handle diverse data types in LAE environments, such as UAVs' CSI, etc.

\noindent \textbf{\textit{2): Pre-processing Module:}}

The \textit{Pre-processing Module} is designed to process raw data from the \textit{Input Module} to ensure compatibility with the LLM. It usually include \textit{Normalization}, \textit{Discretization}, \textit{Domain Transformation}, and \textit{Data selection} process, which ensures that heterogeneous inputs are structured and optimized.

\noindent \textbf{\textit{3): Input encoders:}}

The \textit{Input encoders} are responsible for transforming pre-processed data into numerical representations which the LLM can process. This component typically employs structure such as multilayer perceptron (MLP), transformer block, which is designed based on the features of input data.

\noindent \textbf{\textit{4): LLM Backbone With Finetuning:}}

For the LLM backbone, without loss of generality, we choose GPT-2 with feature dimension $d=768$ as the LLM backbone\cite{gpt2}. It should be noted that, in the proposed method, the GPT-2 backbone can be flexibly replaced with other LLMs, such as Llama and Qwen. The decision of model architecture and scale requires evaluation of the trade-off between computational complexity and performance. As illustrated in Fig.~\ref{figureoverallframework}, GPT-2’s structure relies on stacked transformer decoders. During the training process, only addition, layer normalization layers are fine-tuned for adapting the LLM to the specific task while self-attention and MLP layers are frozen to retain universal knowledge.  

\noindent \textbf{\textit{5): Output decoders:}}

Similar to \textit{Input encoders}, the \textit{Output decoders} are designed to convert the output features of the LLM into the final results for different tasks. They translate the LLM’s internal representations into actionable outputs and ensure that the LLM’s outputs are interpretable and actionable.

\noindent \textbf{\textit{6): Output Module:}}

Finally, the \textit{Output Module} is designed to output results that meet the requirements of different tasks. It may includes the estimated user location, hybrid-field beamformers, etc.

\subsection{Case Study and Simulation Results}

 	\begin{figure*}
	 	\begin{center}
	 		\vspace*{-1mm}\hspace*{-1mm}\includegraphics[width=0.9
    \linewidth]{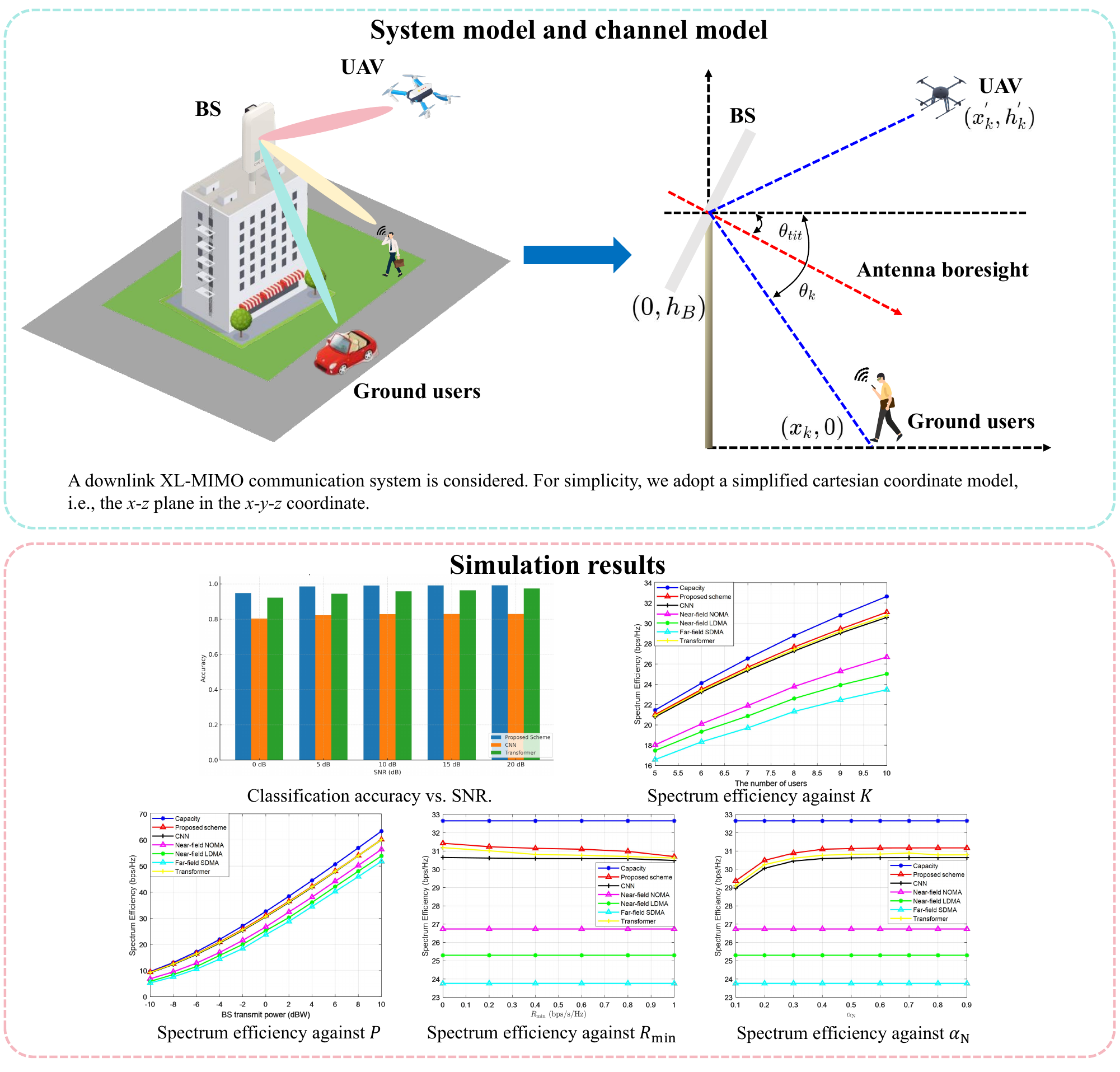}
	 	\end{center}\vspace*{-2mm}	
		\centering
		\caption{Case study and simulation results. A downlink XL-MIMO system is considered, where $N=256$, $h_B=15~\rm{m}$, and $\theta_{tit} = 5$°. The carrier frequency is 30 GHz. The users are randomly distributed, where the range of $x_k$ and $h_k$ are $[0,200~\rm{m}]$ and $[0,30~\rm{m}]$.} 
		\vspace*{-2mm}
		\label{figurecasestudy} 	
	 \end{figure*}

In this subsection, we present specific cases and simulation results to verify the performance of the above LLM-based solutions for near-field communications in LAE. We consider two tasks of the downlink XL-MIMO systems: \textit{distinguishing between far and near-field users} and \textit{multi-user precoding}. 

\noindent \textbf{1): Distinguishing Between Far and Near-Field Users:}

For the user classification task, a consolidated input
$\mathbf{H}=[\mathbf{h}_{1},\mathbf{h}_{2},\cdots,\mathbf{h}_{K}]\in \mathbb{C}^{N \times K}$ is first formed by concatenating the complex channel of the $K$ users. Lastly, the output features of the LLM are converted into the final user classification results by the output projection module.

\noindent \textbf{2): Multi-user Precoding:}

For the multi-user precoding task, its goal is to maximize the spectrum efficiency of multi-user communications. 
It has been previously proven in literature that the optimal downlink beamforming vectors for the spectrum efficiency maximization problem can be obtained as $\mathbf{w}_k^*=\frac{(\mathbf{I}_N+\sum_{k=1}^K\frac{\lambda_k}{\sigma^2}\mathbf{h}_k\mathbf{h}_k^H)^{-1}\mathbf{h}_k}{||(\mathbf{I}_N+\sum_{k=1}^K\frac{\lambda_k}{\sigma^2}\mathbf{h}_k\mathbf{h}_k^H)^{-1}\mathbf{h}_k||_2}$\cite{precnn}. Therefore, we only need to learn $\mathbf{\lambda} = [\lambda_{1},\lambda_{2},\cdots,\lambda_{K}]$ rather than the entire high-dimension matrix $\mathbf{W}$, to obtain the normalized precoding vector. After the output projection module, the $\mathbf{\lambda}$ and power allocation vector $\mathbf{p}=[P_1,P_2,\cdots,P_K]$ can be obtained.

For simulation setups, a training dataset comprising 8,000 samples, a validation dataset comprising 1,000 samples and a testing dataset comprising 1,000 samples are constructed respectively. For the hyper-parameters in network training, we set number of training epoch as 500, the batch size as 100, and learning rate as 0.0001.  All the training and inference of the proposed model is conducted on an NVIDIA GeForce RTX 4090 24GB GPU. Besides, some benchmark comparison schemes are considered, including \textit{Capacity}, \textit{CNN}, \textit{Transformer}, \textit{Near-field NOMA}, \textit{Near-field LDMA}, and \textit{Far-field SDMA}.

As shown in Fig.~\ref{figurecasestudy}, performance comparison of our proposed scheme with the benchmark scheme from multiple perspectives are carried out. According to the simulation results, it can be concluded that our proposed scheme can demonstrate excellent performance under various parameter settings, which verifies the strong robustness and generalization ability of the proposed LLM-based scheme. Specifically, AI-based methods achieve higher
spectrum efficiency than traditional codebook-based methods due to their feature extraction capabilities and increased degrees of freedom in solution space. Furthermore, owing to the increasing size of network, LLM-based method exhibits superior optimization and generalization capabilities and outperforms other deep learning-based methods in terms of performance.

\section{Future Directions and Open Issues}\label{S4}		

In this section, several future research directions for LLM-empowered communications in LAE are discussed.

\subsection{LLM-Empowered Multi-Task Network for LAE}\label{S4.1}

The LAE requires communication systems that can handle multiple tasks in dynamic environments, such as user classification, precoding, channel estimation, and channel prediction. Multi-task LLM offers a unified approach by performing these tasks within a single framework, enhancing efficiency and adaptability in near-field communications. Recently, a LLM-enabled multi-task physical layer network has been proposed in \cite{zheng2024large}, which unifies multiple tasks into a single LLM by carefully designing the multi-task instruction module, input encoders, and output decoders to adapt to the features of different wireless data formats. Future research should focus on integrating additional tasks like interference management, leveraging edge computing, and conducting field tests to ensure robustness in LAE scenarios, paving the way for intelligent low-altitude networks.

\subsection{Multi-Modal LLM for LAE}\label{S4.2}

The advent of multi-modal large language models (MMLLM) represents a pivotal advancement in artificial intelligence, enabling the integration and processing of diverse data types such as text, images, audio, and video\cite{chen2024multi}. This capability is particularly crucial for the LAE, where applications like UAVs and urban air mobility rely on real-time analysis of multiple data modalities for safe and efficient operations. For instance, a MMLLM could allow a drone to simultaneously process visual data for navigation, interpret voice commands, and consult textual flight plans, thereby enhancing its autonomy and situational awareness. Thus, exploring multi-modal LLM for LAE is not only meaningful but essential for advancing the future of low-altitude networking and services.

\subsection{Deployment and Verification of LLM in LAE}

The deployment and verification of LLM in the LAE are critical steps toward realizing practical and efficient low-altitude wireless networks. As LAE applications continue to expand, it comes with unique challenges that must be addressed to ensure successful deployment and reliable performance\cite{zhao2025generative}. For example, LLM requires significant processing power and memory. In LAE, where devices like UAVs have limited onboard computing capabilities, deploying LLMs poses a challenge. Furthermore, devices like UAVs are battery-powered and the computational demands of LLM can significantly impact battery life, necessitating energy-efficient model designs and inference techniques. Besides, integrating LLM into diverse LAE systems requires compatibility with existing hardware, software, and communication protocols.

\section{Conclusions}	
In this article, we conduct an overview of fundamentals, potentials, solutions, and future directions for LLM-empowered near-field communications in LAE. As key technology, physical layer communication architecture, and application scenario for future 6G, the ingenious integration of LLM, near-field communications, and LAE is poised to unlock new possibilities and opportunities for future LAE networks. However, LLM-empowered near-field communications in LAE also faces numerous implementation challenges, necessitating collaborative efforts from both academia and industry to drive its advancement.

	\vspace*{3mm}
 
	\bibliography{IEEEabrv,Reference}

 \begin{IEEEbiographynophoto}
 {Zhuo Xu}
	 is currently an M.S. student in the Department of Electronic Engineering at Tsinghua University, Beijing, China. 
\end{IEEEbiographynophoto}
\vspace{-1cm}

 \begin{IEEEbiographynophoto}
 {Tianyue Zheng}
	 is in the pursuit of a Ph.D.
	degree with the Department
	of Electronic Engineering at
	Tsinghua University, Beijing,
	China. 
\end{IEEEbiographynophoto}
\vspace{-1cm}

\begin{IEEEbiographynophoto}
{Linglong Dai} is a Professor at the Department of Electronic Engineering, Tsinghua University. 
\end{IEEEbiographynophoto}

\end{document}